\documentclass[iop,apj]{emulateapj}
\usepackage{booktabs}

\usepackage{graphicx,epstopdf,hyperref,color,multirow,amsmath}

\newcommand{\Teff}{\mbox{$T_{\mathrm{eff}}$}}

\begin{document}
\title{Machine Learning Regression of stellar effective temperatures in the second $Gaia$ Data Release }

   \author{Yu Bai\altaffilmark{1}}
   \author{JiFeng Liu\altaffilmark{1,2}}
   \author{ZhongRui Bai\altaffilmark{1}}
   \author{Song Wang\altaffilmark{1}}
   \author{DongWei Fan\altaffilmark{3}}

\affiliation{$^1$ Key Laboratory of Optical Astronomy, National Astronomical Observatories, Chinese Academy of Sciences,
       20A Datun Road, Chaoyang Distict, Beijing 100012, China; ybai@nao.cas.cn}
\affiliation{$^2$ College of Astronomy and Space Sciences, University of Chinese Academy of Sciences, Beijing 100049, China}
\affiliation{$^3$ National Astronomical Observatories, Chinese Academy of Sciences,
       20A Datun Road, Chaoyang Distict, Beijing 100012, China}

\begin{abstract}
This paper reports on the application of the supervised machine-learning algorithm to the stellar effective temperature
regression for the second $Gaia$ data release, based on the combination of the stars in four spectroscopic surveys:
Large Sky Area Multi-Object Fiber Spectroscopic Telescope,  Sloan Extension for Galactic Understanding and Exploration,
the Apache Point Observatory Galactic Evolution Experiment and the RAdial Velocity Extension. This combination,
about four million stars, enables
us to construct one of the largest training sample for the regression,
and further predict reliable stellar temperatures with a root-mean-squared error
of 191 K. This result is more precise than that given by $Gaia$ second data release that is based on about sixty thousands stars. 
After a series of data cleaning processes,
the input features that feed the regressor are carefully selected from the $Gaia$ parameters, including the colors, the 3D position
and the proper motion.
These $Gaia$ parameters is used to predict effective temperatures for 132,739,323 valid stars in the second $Gaia$ data release.
We also present a new method for blind tests and a test for external regression without additional data.
The machine-learning algorithm fed with the parameters only in one catalog provides us an effective approach to maximize
sample size for prediction, and this methodology has a wide application prospect in future studies of astrophysics.

\end{abstract}


\section{Introduction}
\label{sec:intro}
The ESA space mission $Gaia$ is performing an all-sky astrometric, photometric and radial velocity survey at
optical wavelength \citep{Gaia16}. The main objective of the $Gaia$ mission is to survey more than one billion
stars, in order to understand the structure, formation, and evolution of our Galaxy. The second data release
($Gaia$ DR2; \citealt{Gaia18}) includes a total of 1.69 billion sources with $G$-band photometry based on 22
months of observations. Of these, 1.38 billion sources also have the integrated fluxes from the BP and RP
spectrophotometers, which span 3300$-$6800 {\AA} and 6400$-$10500 {\AA}, respectively.

These three broad photometric bands have been used to infer stellar effective temperatures ({\Teff}), for all
sources brighter than $G$ $=$ 17 mag with {\Teff} in the range 3000$-$10,000 K \citep{Andrae18}. A machine
learning algorithm, random forest (RF), has been applied to regress {\Teff}. The training data of the algorithm
is a combination of five spectrum- or photometry-based catalogs with a total 65,200 stars.
A typical accuracy of the regression is 324 K that is estimated from 50\% hold-out validation, and no blind
test is performed to quantify the performance of the regression and to avoid overfitting.

However, decoupling stellar temperatures and interstellar extinction is a complex problem, and more parameters than
two colors is required to regress temperatures with good accuracy \citep{Bai19}.
Moreover, diversity of a sample in a parameter space has been proved to be an influential aspect, and has strong impact on
overall performance of machine learning (\citealt{Wang09a}, \citealt{Wang09b}).
The small size of training set in \citet{Andrae18} could limit the diversity of the stellar sample and further cause
regressed {\Teff} having high systematic deviation (e.g., \citealt{Pelisoli19}; \citealt{Sahlholdt19}).

The availability of spectrum-based stellar parameters for large numbers is now possible thanks to the observations
of large Galactic spectral surveys.
Large Sky Area Multi-Object Fiber Spectroscopic Telescope (LAMOST; \citealt{Luo15}) data release 5 (DR5)
was available to domestic users in December of 2017, which includes over eight millions observations of stars \footnote{See http://dr5.lamost.org/.}.
This archive data after six years' accumulation is a treasure for various studies. One of the catalog mounted on the
archive is A, F, G and K type stars catalog, in which the stellar parameters, {\Teff}, log$g$ and [Fe/H] are determined
by the LAMOST stellar parameter pipeline \citep{Wu14}.

Another large survey is Sloan Extension for Galactic Understanding and Exploration
(SEGUE; \citealt{Yanny09}). The spectra are processed through the SEGUE Stellar Parameter Pipeline
(SSPP; \citealt{Allende08,Lee08a}; \citealt{Lee08b,Smolinski11}), which uses a number of methods to derive accurate
estimates of stellar parameters, {\Teff}, log$g$, [Fe/H], [$\alpha$/Fe] and [C/Fe].

Different from the upper two surveys that are in optical band, the Apache Point Observatory Galactic Evolution Experiment (APOGEE),
as one of the programs in both SDSS-III and SDSS-IV, has collected high-resolution ($R$ $\sim$ 22,500) high signal-to-noise (S/N $>$ 100)
near-infrared (1.51$-$1.71 $\mu$m) spectra of 277,000 stars (data release 15) across the Milky Way \citep{Majewski17}.
These stars are dominated by red giants selected from the Two Micron All Sky Survey.
Their stellar parameters and chemical abundances are estimated by the APOGEE Stellar Parameters and Chemical
Abundances Pipeline (ASPCAP; \citealt{Garcia16}).

These surveys aim mainly at stars located in the north hemisphere, while the RAdial Velocity Extension (RAVE) covers
the south sky.
It is designed to provide stellar parameters to complement missions that focus on obtaining
radial velocities to study the motions of stars in the Milky Way¡¯s thin and thick disk and stellar halo \citep{Steinmetz06}.
Its pipeline processes the RAVE spectra and derives estimates of \Teff, log $g$, and [Fe/H] \citep{Kunder17}.

The large amount of spectroscopic data in these four catalogs provides us an opportunity to apply machine learning technology
to regress {\Teff} effectively.
In Section \ref{sec:meth}, we present validation samples and a method of data cleaning. Various input parameters are also
explored to regress temperatures in the section. We apply the regressor and present a revised
version of {\Teff} catalog for $Gaia$ DR2 in Section \ref{sec:res}. Blind tests and external regression tests are also provided.
A discussion is given in Section \ref{sec:sum}.

\section{Methodology}
\label{sec:meth}
\subsection{Validation Samples}
The A, F, G and K type stars catalog of LAMOST DR5 includes the estimates of the stellar {\Teff} with the application of a
correlation function interpolation \citep{Du12} and Universit\'{e} de Lyon spectroscopic analysis software \citep{Koleva09}.
These two approaches are based on distribution and morphology of absorption lines in normalized stellar spectra, independent
from Galactic extinction.  The temperatures are in the rang of 3460 $<$ {\Teff} $<$ 8500 K with the uncertainty of $\sim$110 K
\citep{Gao15}. We extract 4,340,931 unique stars in the catalog, and cross match them to $Gaia$ DR2 with a radius of 2 arcseconds,
which yields 4,249,013 stars.

For SEGUE survey, we adopt {\Teff} estimated with the SSPP that is also based on distribution and morphology of stellar
absorption lines.
The temperatures range from 4000 $<$ {\Teff} $<$ 9710 K with the typical uncertainty of $\sim$180 K. We perform a cross
match with $Gaia$ DR2, and obtain 1,037,433 stars.

The {\Teff} of APOGEE stars is estimated by ASPCAP, which searches a multi-dimensional grid for the best-matching synthetic
spectrum \citep{Meszaros13}. The temperatures are in the range of 3550 $<$ {\Teff} $<$ 8200 K, with the typical uncertainty
of $\sim$100 K. We cross match these stars with $Gaia$ DR2, and obtain 275,019 stars.

The pipeline of RAVE is based on the combination of the MATrix Inversion for Spectral SynthEsis (MATISSE; \citealt{Recio06})
algorithm and the DEcision tree alGorithm for AStrophysics (DEGAS; \citealt{Bijaoui12}). This pipeline is valid for
stars with temperatures between 4000 K and 8000 K. The estimated errors in {\Teff} is approximately 250 K, and $\sim$100 K
for spectra with S/N $\sim$ 50 \citep{Kunder17}. The cross match
with $Gaia$ DR2 yields 518,812 stars.

We here only adopt the {\Teff} from spectroscopic surveys, since their stellar parameters are highly
reliable \citep{Mathur17}, compared to photometric surveys, e.g., Kepler Input Catalog. As a result, there are 6,080,277
$Gaia$ matched stars in the four catalogs.

\subsection{Data Cleaning}

\citet{Andrae18} applied various filters to remove bad data, and some of them are also adopted in our data cleaning processes.
We remove the samples with $\varpi$ $\leq$ 0 or $\sigma_{\varpi}/\varpi >$ 0.2. The samples with the high or negative
relative uncertainties of the parallaxes may suffer large bias in the distance measurements \citep{Luri18}, or could include large
fraction of non-stellar objects \citep{Bai18}. We also exclude the samples with $\sigma(\Teff)/\Teff >$ 0.05 to remove
inaccurate estimates.

\begin{figure}
   \centering
   \includegraphics[width=0.45\textwidth]{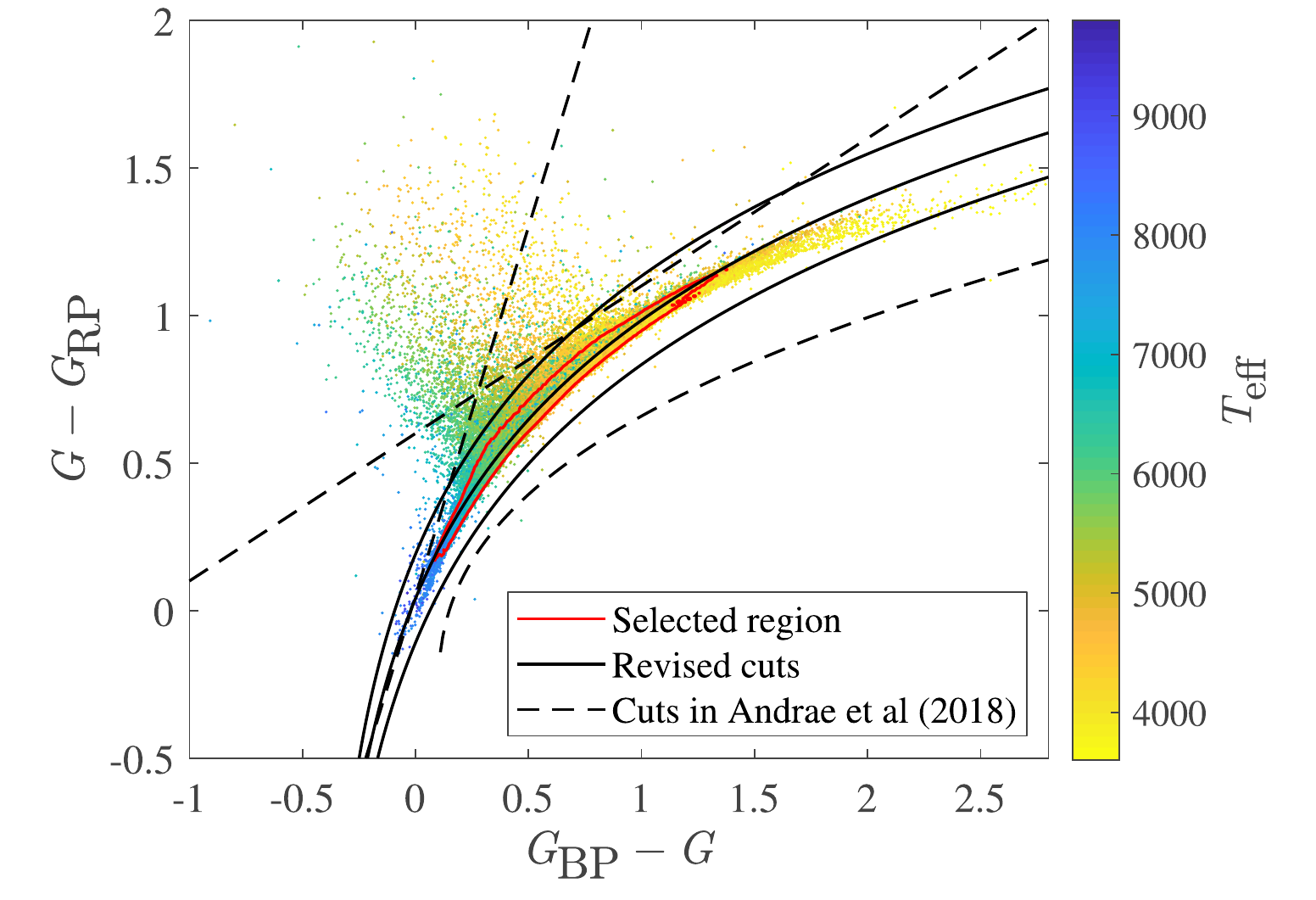}
   \caption{Color-color diagram for $Gaia$ stars matched with four catalogs. Dashed lines are
            quality cuts in \citet{Andrae18}. Solid black lines are our revised quality cuts.
            The red polygon indicates the region with number densities higher than 150 per 0.01 mag$^2$.
            The color bar stands for the {\Teff}. For clarity, we randomly select and plot one tenth of
            the sample.
   \label{CCD}}
\end{figure}

We plot color-color diagram in Figure \ref{CCD}, and select the region with number densities higher than 150 per 0.01 mag$^2$.
A logarithmic function is used to fit the colors of the sample in this region.
The best fit function is $G - G_\textrm{RP}$ = 1.79$\cdot$log$_{10}$($\textit{G}_\textrm{BP} - G$+0.42)+0.71.
We shift the function with $\pm$ 0.15 mag to select the samples with good photometry. This good-quality region is marked with
the black solid lines in Figure \ref{CCD}.
The region defined by the logarithmic function shows a better consistency with the stellar locus than the cuts in \citet{Andrae18}.
As a result, the training sample contains 3,810,143 stars.

\begin{figure}
   \centering
   \includegraphics[width=0.45\textwidth]{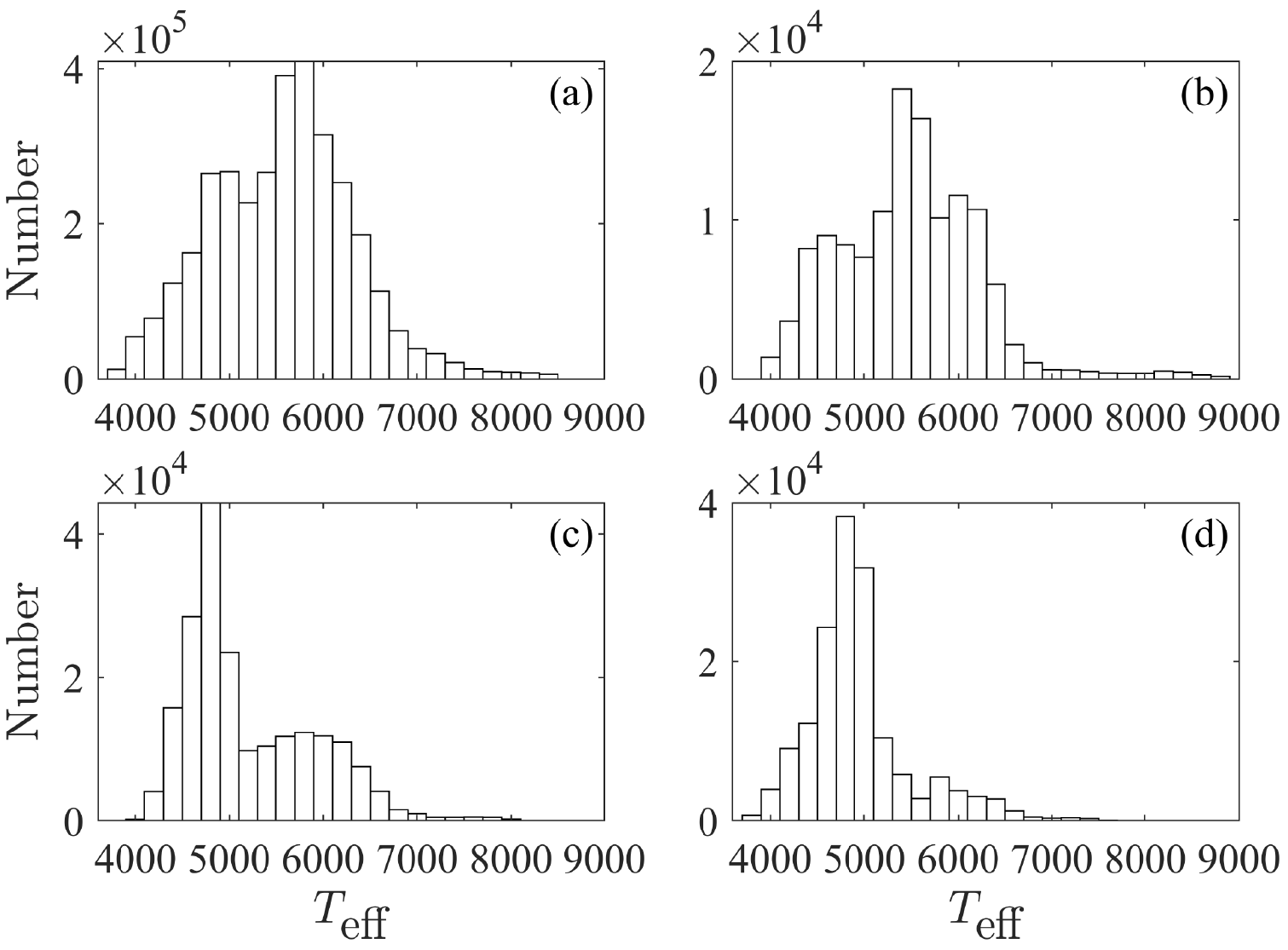}
   \caption{Distributions of the literature {\Teff} in (a) LAMOST, (b) SSPP, (c) RAVE and (d) APOGEE.
   \label{hist}}
\end{figure}

The {\Teff} distribution of these stars is shown in Figure \ref{hist}, which is inhomogeneous. We give the impact of this
on the prediction for $Gaia$ DR2 in Section \ref{sec:res}.
The training sample is dominated by F, G, and K stars with {\Teff} $\sim$ 5000-6000 K, different from the
distribution of the training sample in \citet{Andrae18} that concentrates in five specific temperatures.

\begin{figure}
   \centering
   \includegraphics[width=0.5\textwidth]{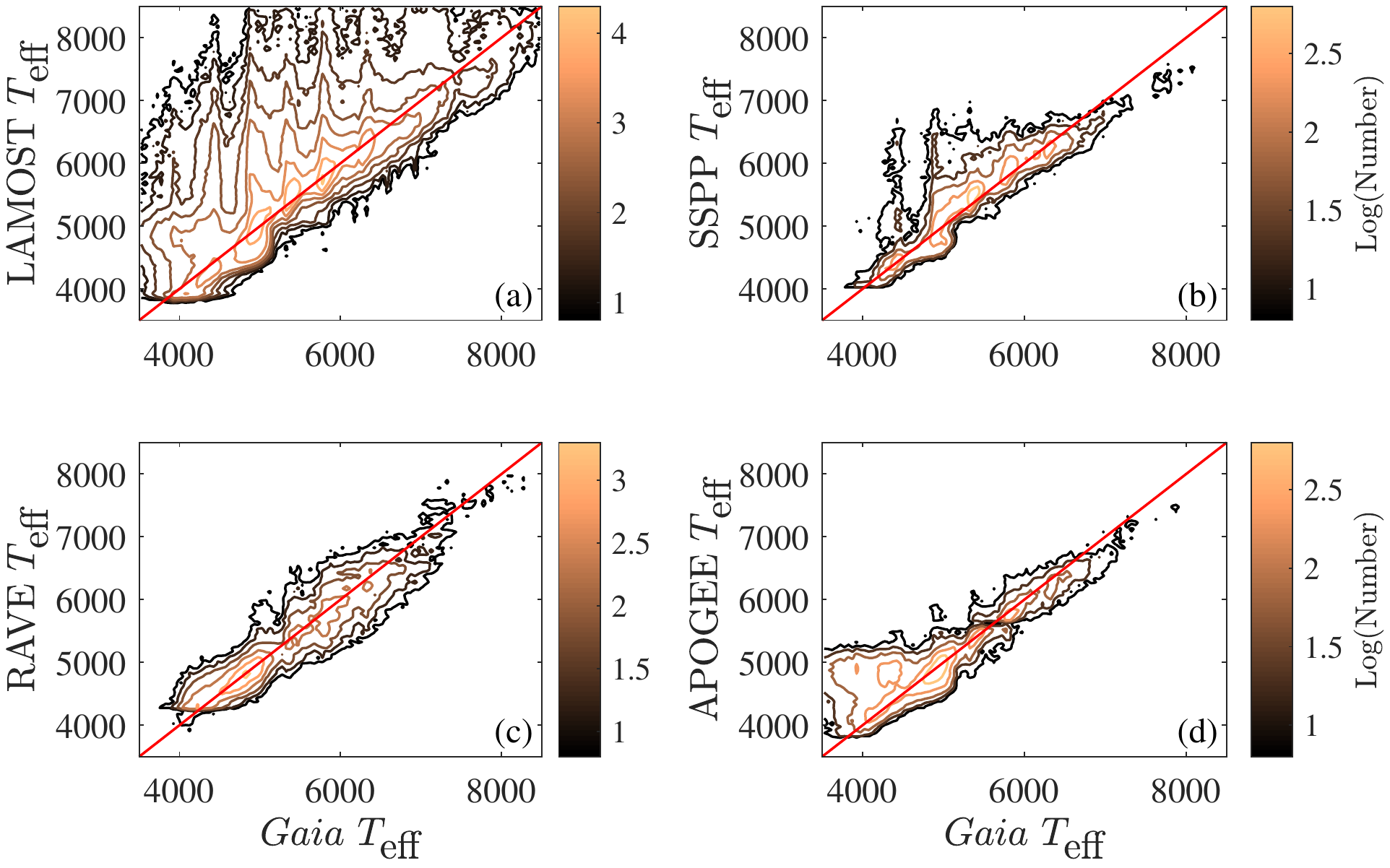}
   \caption{Density contours of one-to-one correlations between the {\Teff} in $Gaia$ DR2 and in
            (a) LAMOST, (b) SSPP, (c) RAVE and (d) APOGEE.
   \label{OneOne}}
\end{figure}

We present the differences between the {\Teff} in $Gaia$ DR2 and the literature estimates in Figure \ref{OneOne}.
Some vertical concentrated regions are shown in LAMOST and SSPP panels. The stars in these regions have similar
temperatures in $Gaia$ DR2, but have different estimates in the spectrum-based catalogs.
This implies that the temperatures given by $Gaia$ DR2 are probably still coupled with Galactic extinction, since the regressor was
built with two colors from a small size sample. These two colors couldn't provide enough information to
decouple the temperatures from the extinction \citep{Davenport14}.

\subsection{Input Parameters}

The {\Teff} in $Gaia$ DR2 was determined using two colors in $Gaia$ photometric bands, $G - G_\textrm{RP}$ and
$G_\textrm{BP} - G$ \citep{Andrae18}.  They tested randomised trees, support vector machine (SVM) and Gaussian processes.
The algorithm of randomised trees showed very fast learning, and its results are as good as the other two algorithms.

We also adopt the random forest algorithm (RF; \citealt{Breiman01}) to build the regressor, but try different combinations of input
parameters. The working theory of the RF is that it builds an ensemble of unpruned decision trees and merges them
together to obtain a more accurate and stable prediction. The algorithm consists of many decision trees, and it outputs the
class that is the mode of the class output by individual trees. The RF is often used when we have very large training
data sets and a very large number of input variables. One big advantage of RF is fast learning from a very large number of
data. 

We here apply the 10 folded cross validations to test the performance of the regression, rather than
the 50\% hold validation that is used in \citet{Andrae18}. The cross validation partitions the sample into ten randomly
chosen folds of roughly equal size. One fold is used to validate the regression that is trained using the remaining folds.
This process is repeated ten times such that each fold is used exactly once for validation. The 10 folded cross validation
can provide an overall assessment of the regression.

\begin{figure}
   \centering
   \includegraphics[width=0.5\textwidth]{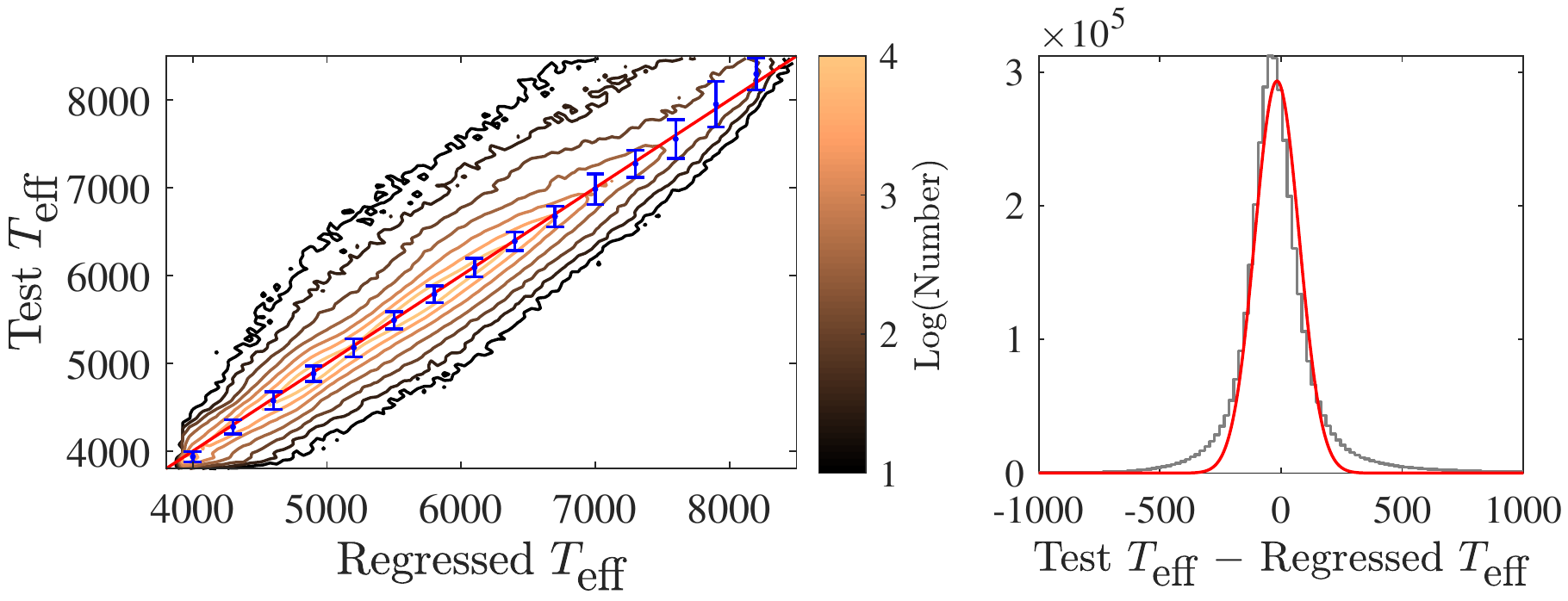}
   \caption{One-to-one correlation of the cross validation (left panel). The blue error bars stand for
            the Gaussian fit in the bin of regressed {\Teff}. The centers of the bars are the Gaussian
            centers, and the lengths of the bars are the Gaussian standard deviations. The color bar is
            the density contour in the log scale. Gaussian fit (red) of the total residual
            (black) is shown in the right panel.
   \label{CV}}
\end{figure}

The root-mean-squared error (RMSE) is adopted to stand for the performance of the regressors (Table \ref{Table1}). We find that
the regressor of eight input parameters, $l$, $b$, $\varpi$, $\Delta\varpi$, $\mu_{\alpha}$, $\mu_{\delta}$,
BP $-$ $G$ and $G$ $-$ RP, shows the best performance with RMSE of 191 K, while the regressor that is constructed
with only two colors is the worst. 
The one-to-one correlation of the best regression is shown in the left panel in Figure \ref{CV}.
We bin the regressed {\Teff} with a step size of 100 K and fit the distribution of the corresponding test {\Teff} with a
Gaussian function (the blue error bars in Figure \ref{CV}), in order to estimate the uncertainty of the regression for different temperatures.
The Gaussian fit to the total residuals is shown in the right panel, and the fitted offset ($\mu$) and the standard deviation ($\sigma$) are
listed in Table \ref{Table3}.

\begin{deluxetable}{rl}
\tablecaption{RMSEs of Different Input Parameters \label{Table1}}
\tablehead{ Parameters & RMSE (K)
           }
\startdata
BP$^*$ $-$ $G$, $G$ $-$ RP$^*$        & 407\\
$G$, BP, RP                 & 393\\
$\alpha$, $\Delta\alpha$, $\delta$, $\Delta\delta$, $\varpi$, $\Delta\varpi$, $\mu_{\alpha}$, & \multirow{2}{*}{227}\\
$\Delta\mu_{\alpha}$, $\mu_{\delta}$, $\Delta\mu_{\delta}$, $G$, BP, RP &  \\
$\alpha$, $\Delta\alpha$, $\delta$, $\Delta\delta$, $\varpi$, & \multirow{2}{*}{226}\\
$\Delta\varpi$, $G$, BP, RP                 &    \\
$\alpha$, $\Delta\alpha$, $\delta$, $\Delta\delta$, $\varpi$, $\Delta\varpi$, $\mu_{\alpha}$, $\Delta\mu_{\alpha}$, & \multirow{2}{*}{198}\\
$\mu_{\delta}$, $\Delta\mu_{\delta}$, BP $-$ $G$, $G$ $-$ RP & \\
$\alpha$, $\delta$, $\varpi$, $\Delta\varpi$, BP $-$ $G$, $G$ $-$ RP & 196 \\
$\alpha$, $\Delta\alpha$, $\delta$, $\Delta\delta$, $\varpi$, $\Delta\varpi$, &\multirow{2}{*}{194}\\
BP $-$ $G$, $G$ $-$ RP         &    \\
$\alpha$, $\delta$, $\mu_{\alpha}$, $\mu_{\delta}$, $\varpi$, $\Delta\varpi$, & \multirow{2}{*}{193} \\
BP $-$ $G$, $G$ $-$ RP & \\
$l$, $b$, $\varpi$, $\Delta\varpi$, BP $-$ $G$, $G$ $-$ RP & 192 \\
$l$, $b$, $\varpi$, $\Delta\varpi$, $\mu_{\alpha}$, $\mu_{\delta}$, & \multirow{2}{*}{191}\\
BP $-$ $G$, $G$ $-$ RP &
\enddata
\tablecomments{BP and RP are the photometry in the bands of $\textit{G}_\textrm{BP}$ and $\it{G}_\textrm{RP}$.
}
\end{deluxetable}

\section{Result}
\label{sec:res}
We now use the criteria below to select the samples in $Gaia$ DR2, which yields 132,739,323 stars.
\begin{align*}
&\Delta{\Teff}/{\Teff} < 0.05, \\
&0 < \Delta\varpi/\varpi < 0.2, \\
&G - G_\textrm{RP} \leq 1.79{\cdot}\textrm{log}_{10}(\textit{G}_\textrm{BP} - G+0.42)+0.71+0.15, \\
&G - G_\textrm{RP} \geq 1.79{\cdot}\textrm{log}_{10}(\textit{G}_\textrm{BP} - G+0.42)+0.71-0.15.
\end{align*}
The algorithm of RF constructed with eight input parameters is applied to regress their {\Teff}, and
the result is listed in Table \ref{Table2}.

\begin{figure}
   \centering
   \includegraphics[width=0.5\textwidth]{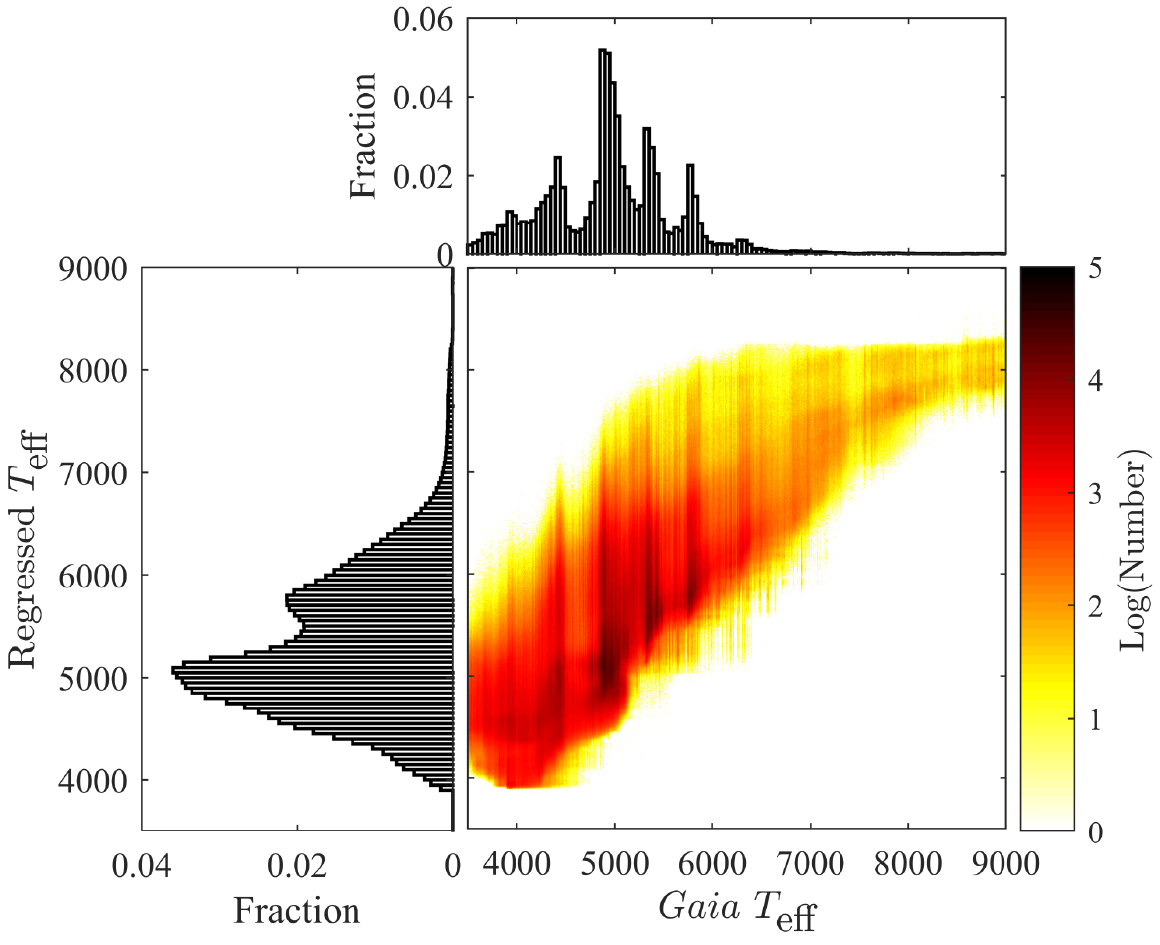}
   \caption{Density map of {\Teff} in $Gaia$ DR2 vs. regressed {\Teff}. Normalized histograms
            of the {\Teff} distributions are plotted in the top and left panels.
   \label{OneOneGaia}}
\end{figure}

The size of the catalog is a little smaller than that in \citet{Andrae18}, since we use more strict criteria.
We compare our results with {\Teff} in $Gaia$ DR2 in Figure \ref{OneOneGaia}. The $Gaia$ {\Teff} is concentrated
in some specific temperatures, 4000 K, 4500 K, 5000 K, 5500 K and 6000 K. These temperatures are consisted with the
peaks in the distribution of the training set of the regressor. The inhomogeneous training set yielded output
with similar distribution (see Fig. 5 and Fig. 18 in \citealt{Andrae18}).
Our {\Teff} distribution concentrates in two much broad peaks, 5000 K and 6000 K, implying better homogeneousness.

\begin{deluxetable}{cl}
\tablecaption{Results of our regression in $Gaia$ DR2 \label{Table2}}
\tablehead{ Source ID & Regressed {\Teff}
           }
\startdata
2448780173659609728 & 5128 $\pm$ 634\\
2448781208748235648 & 5463 $\pm$ 69\\
2448689605685695488 & 5984 $\pm$ 91\\
2448689777484387072 & 4333 $\pm$ 396\\
2448783991887042176 & 4166 $\pm$ 65\\
2448690258520723712 & 5062 $\pm$ 55\\
2448690327240200576 & 5846 $\pm$ 89\\
2448689811844125184 & 4328 $\pm$ 385\\
2448784953959717376 & 5382 $\pm$ 58\\
2448783991887042048 & 4888 $\pm$ 539
\enddata
\tablecomments{This table is available in its entirety in machine-readable form.
}
\end{deluxetable}

\subsection{Blind Tests}
\label{sec:blin}

Blind test is effective method to measure performance of a machine learning classifier or regressor \citep{Bai19}.
It evaluates prediction accuracy with data that are not in the training set, and provides validation that a regressor is
working sufficiently to output reliable results.

\begin{figure}
   \centering
   \includegraphics[width=0.5\textwidth]{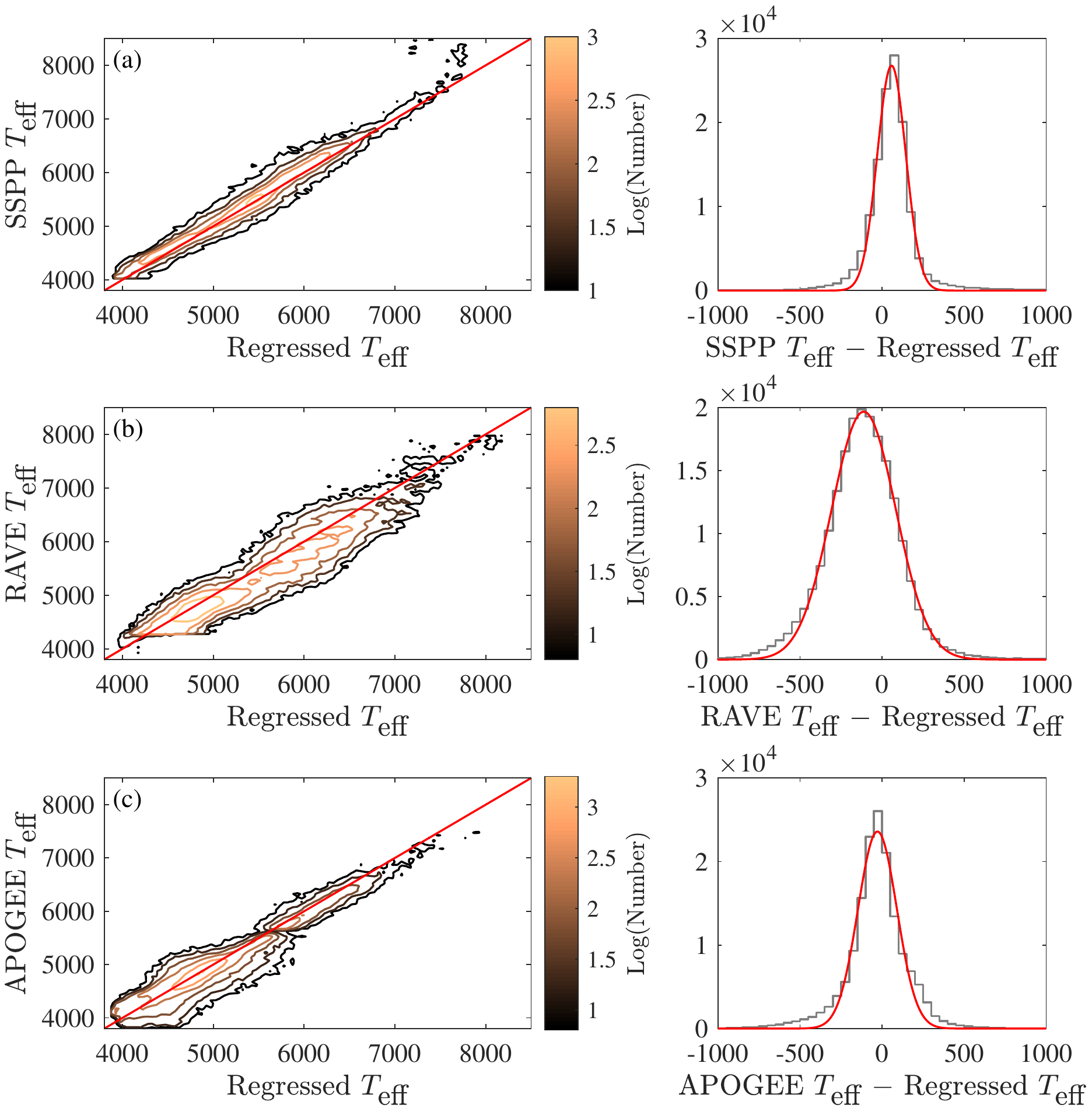}
   \caption{Density contours of one-to-one correlations (left collum) and Gaussian fits of the total residual
            (right collum). Three catalogs are used for training and the rest one for a blind test.
            The test catalogs are: (a) SSPP, (b) RAVE and (c) APOGEE.
   \label{BlindT}}
\end{figure}

In order to apply blind tests, we train sub-regressors with eight input parameters in three catalogs, and use the
forth catalog to test these sub-regressors.
The LAMOST DR5 is always included in the training set, since it accounts for 87\% of the stars in our training set.
We omit the testing stars that located outside the parameter spaces of the sub-regressors to avoid external regression.
We present the results of the blind tests in Figure \ref{BlindT}, and list the parameters of the Gaussian fit to the
total residuals in Table \ref{Table3}.

The blind tests show that the offsets of the total residuals are below 112 K, and the standard deviations are less than 200 K.
\citet{Lee15} has applied the SSPP to LAMOST stars and compared the results to those from RAVE and APOGEE catalogs.
The offsets of {\Teff} between different pipelines are from 36 to 73 K, and the standard deviations are from 79 to 172 K.
This indicates that our regressor can output the stellar temperatures at similar accuracy to the results of spectrum-based
pipelines.

\begin{deluxetable}{lrrc}
\tablecaption{Results of cross validation and blind tests \label{Table3}}
\tablehead{& $\mu$ & $\sigma$ & RMSE \\
           &   (K)  &   (K)    &  (K)
           }
\startdata
Cross Validation       & $-$17  $\pm$ 1     &  91 $\pm$ 1 & 191\\
SSPP                   &    58  $\pm$ 2     &  87 $\pm$ 2 & 179\\
RAVE                   & $-$112 $\pm$ 4     & 196 $\pm$ 4 & 260\\
APOGEE                 & $-$28  $\pm$ 3     & 119 $\pm$ 3 & 191
\enddata
\end{deluxetable}

\subsection{External Regression}
\label{sec:blin}
In order to test the stars that are located outside our criteria, we adopt the sub-regressors trained with three catalogs and
use the stars in the forth catalog to apply external regression. The stars are divided into two subclasses, located outside the
quality cuts in Figure \ref{CCD}, and with 0.2 $< \Delta\varpi/\varpi <$ 0.4 (Table \ref{Table4}).

\begin{figure}
   \centering
   \includegraphics[width=0.5\textwidth]{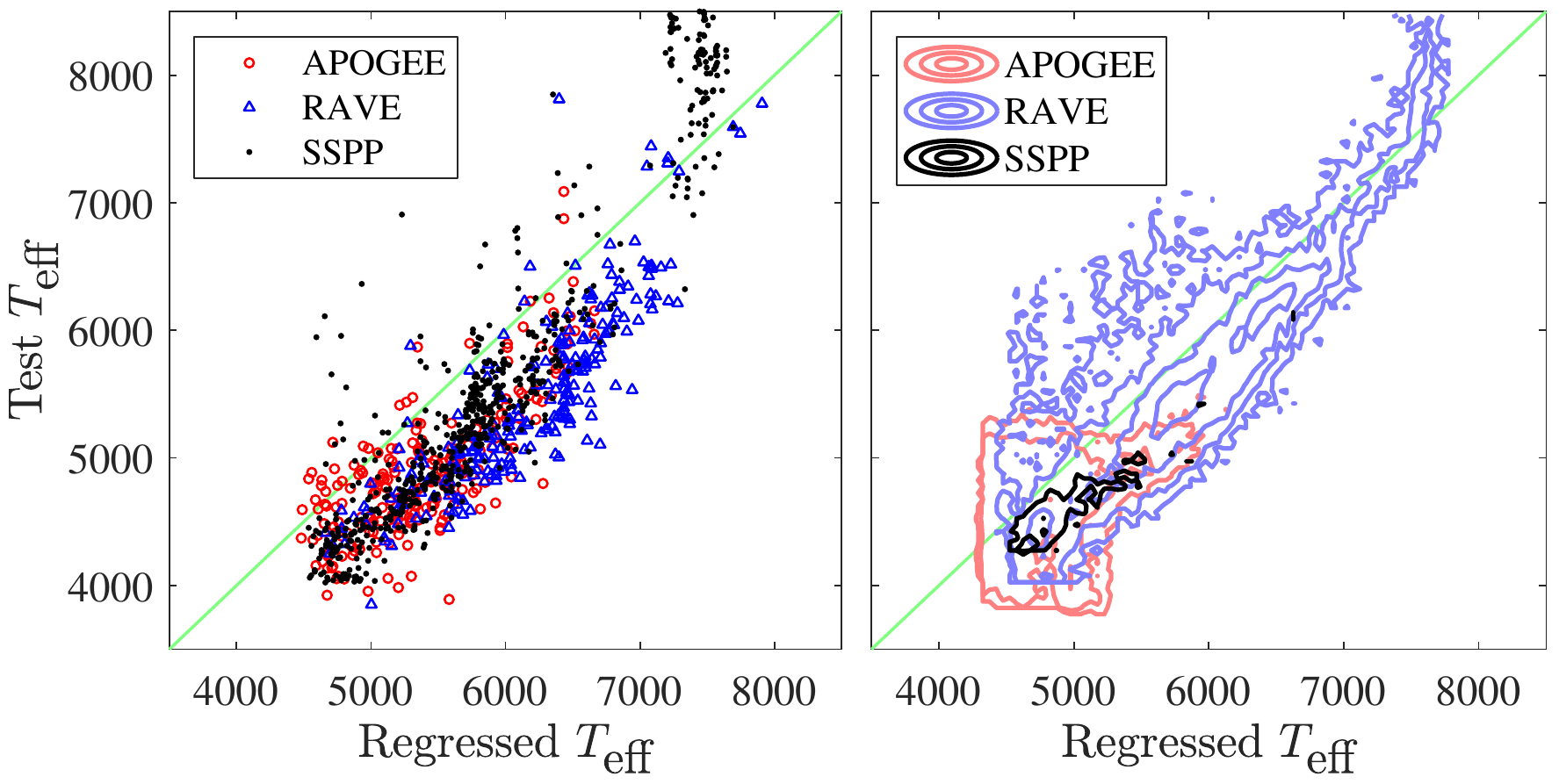}
   \caption{The result of external regression.
            Left panel: the stars that located outside the quality cuts
            in Figure \ref{CCD}.
            Right panel: the stars with 0.2 $< \Delta\varpi/\varpi <$ 0.4.
   \label{ExReg}}
\end{figure}

The result is shown in Figure \ref{ExReg}. The {\Teff} is systematically overestimated for the first testing subclass, and their RMSEs
are twice larger than those of the blind tests. The photometry that feed to the sub-regressors is probably worse than the photometry that
located inside the quality cuts, and the sub-regressors could not predict {\Teff} with good accuracy.

For the second subclass, most of the stellar temperatures are also overestimated, since a large parallax relative uncertainty may
refer to a complex transformation to determine a distance \citep{Bailer18}. Such a transformation may bring noise to the sub-regressors
and results in bad performances.

We don't test the regression with {\Teff} outside the training label range of 3700$-$9700 K because of the inability of RF to
extrapolate. \citet{Andrae18} fed their regressor with stars that have {\Teff} outside the training interval, and those stars
were assigned temperatures inside the training interval.

Therefore, it is suggested that all the criteria should be applied before regression in order to select good samples and further
produce reliable {\Teff}.

\begin{deluxetable}{lrrc}
\tablecaption{Numbers and RMSEs of external tests \label{Table4}}
\tablehead{ Catalog & First subclass & Second subclass
           }
\startdata
SSPP                   & 666 (383)& 74,015 (184) \\
RAVE                   & 247 (510)&  1,245 (314) \\
APOGEE                 & 225 (406)& 22,444 (315)
\enddata
\tablecomments{The numbers in the brackets are RMSEs in the unit of Kelvin.
}
\end{deluxetable}

\section{Discussion}
\label{sec:sum}
In this work, we have attempted to regress the effective temperatures for 132,739,323 stars in $Gaia$ DR2 using machine
learning algorithm. The regressor is trained with about four million stars in LAMOST, SSPP, RAVE and APOGEE catalogs,
one of the largest training sample ever used for machine learning in astrophysics.
We have tried several combinations of input parameters, and have applied cross validation to test the performances.
The regressor with the smallest RMSE is built with $l$, $b$, $\varpi$, $\Delta\varpi$, $\mu_{\alpha}$, $\mu_{\delta}$,
BP $-$ $G$ and $G$ $-$ RP. The cross validation indicates that the typical accuracy of the regression is 191 K.
In order to examine the performance of the regressor, we use the majority of the training set to build three sub-regressors,
and apply the rest small fraction for blind testing.  The testing results show similar performance to some spectrum-based
piplines. In this section we would like to discuss the processes that haven't been used in other machine-learning studies.

\subsection{Feature Selection}
In machine learning technology, feature selection is a process of selecting features in the data that are most useful or
most relevant for the problem. The problem in the paper is regressing {\Teff} with parameters in $Gaia$ DR2.
We adopt the RMSE to indicate the relevance of the problem for the different subsets of the input parameters.

One of the most popular parameters is a stellar color, since a stellar temperature could be roughly described by a color.
However, this description suffers from temperature-extinction coupling.
When we try to use $Gaia$ colors or magnitudes to regress {\Teff}, the performance is bad.
This implies that the color parameters are relevant to our problem, but the problem couldn't be fully described by these colors.
The additional input parameters are required to provide information about the Galactic interstellar extinction.

\begin{figure}
   \centering
   \includegraphics[width=0.45\textwidth]{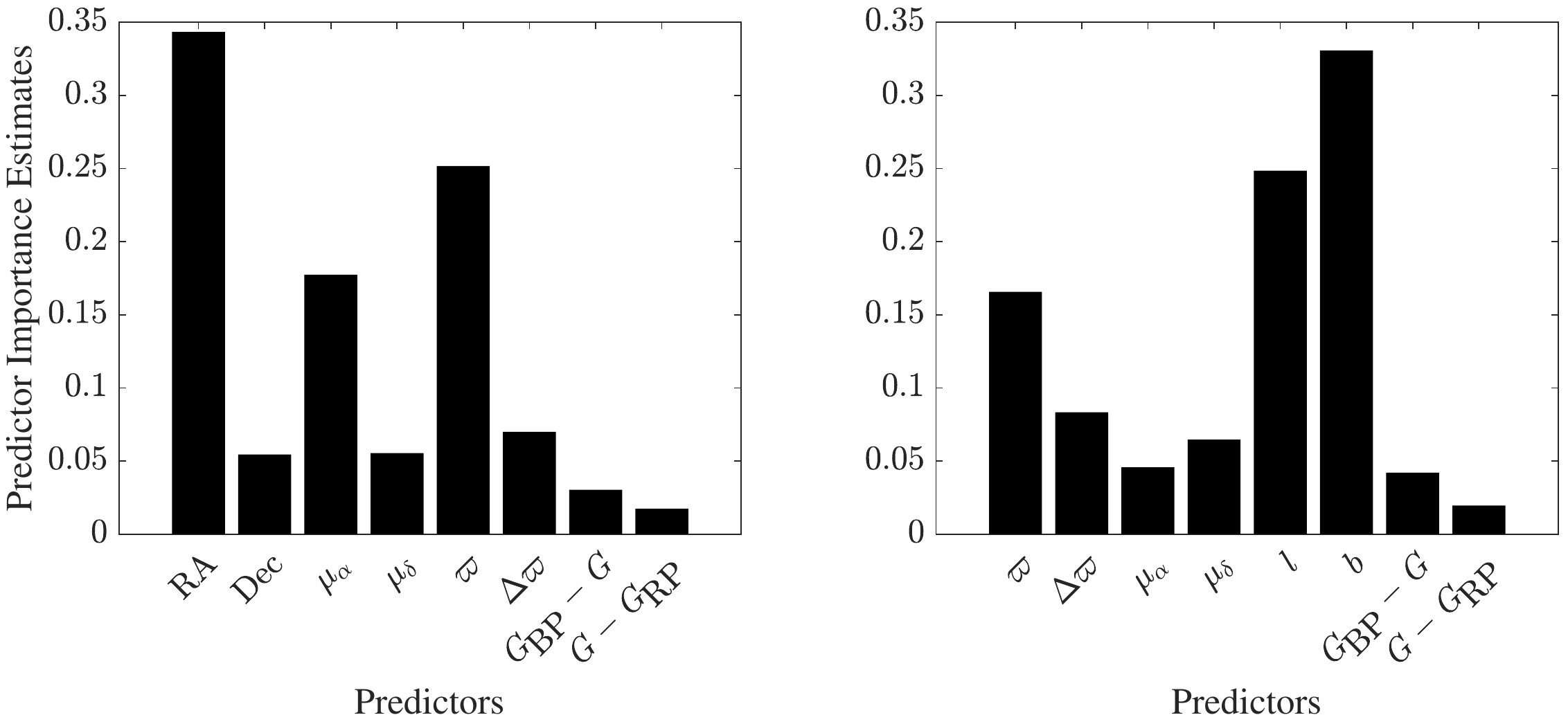}
   \caption{The importance estimates of two different subsets of input parameters.
            The final adopted regressor is built with the parameters in the right panel.
   \label{Imp}}
\end{figure}

Many works have been done to draw the 3D dusty map of the Milky Way (eg. \citealt{Green18}). The extinction value
is a function of the stellar location. When we add $\alpha$, $\delta$ and parallax to the parameter subsets,
the performance becomes better. The RMSE is slightly smaller for the regressor with $l$ and $b$ input than $\alpha$ and
$\delta$ input, probably due to the transformation between equatorial and galactic coordinates.
The algorithm need to find this potential transformation when build the regressor with $\alpha$ and $\delta$, which
may add additional noise and result in a larger RMSE. When we use $l$ and $b$ instead of $\alpha$ and $\delta$
to build the regressor, $l$ and $b$ become the most two important parameters (Figure \ref{Imp}).
It implies that the information on Galactic extinction plays an important role in the {\Teff} regression.

The proper motion can also improve the performance of the regressor, and its importance is higher than those of
$Gaia$ colors (Figure \ref{Imp}), implying that its more relevant than colors in our {\Teff} regressing process.
The proper motion could provide assistant information on stellar distance
statistically, based on the fact that the systematic errors in distance would result in the correlations between
the measured $U$, $V$, and $W$ velocity components \citep{Schonrich12,Wang16}.
This implies that when we add the proper motion to the parameter subsets, the parallax could give more information
about the reliability of the stellar distance.


\subsection{Blind Tests for Sub-regressors}
The training set for the regressor is dominated by the stars in the LAMOST catalog, over 87\%, and the other three
catalogs comprise $\sim$4\% of SSPP, $\sim$5\% of RAVE and $\sim$4\% of APOGEE.
We build three sub-regressors with a combination of three catalogs that are $\sim$95\% of the training set, in order to apply blind tests
and further to avoid potential overfitting.
Each one of the sub-regressor could be used to predict the {\Teff} for the stars in $Gaia$ DR2, while we use all four
catalogs to train the final regressor in order to maximize the performance.

It has been proved that the performance of the regressor can be increased by adding more data to the training set \citep{Banko01}.
However, \citet{Pilaszy09} argued that more data not always help to improve the performance, and only good data can rather
than noisy data.
The consistency in the results of our blind tests indicates that the stars in SSPP, RAVE and APOGEE carry more signal than
noise into the training set, and using four catalogs rather than three could raise the performance.

On one hand, the systematic error of the regression is mainly from the biases among different input catalogs (Figure \ref{hist}).
Such error doesn't decrease, when we add more data to build the regressor.
The residuals in Figure \ref{BlindT} show that the biases are constrained to
$\mu < 112$ and $\sigma <$ 200 for four spectrum-based catalogs. On the other hand, performing a regression on the
combination of four catalogs can be just as biased as performing the regression on three of them.
In these cases, it can be reasonable to use an averaging scheme, when there is enough samples in every bin of the grid.
However, the training stars are dominated by F, G and K stars, and the sample sizes of high and low mass aren't enough to
smooth the fluctuation in the bins. We would take advantage of the averaging scheme to train a regressor more effectively,
with the help of $Gaia$ DR3 (next year) and LAMOST DR6 plus early version of DR7 (more than ten million spectra in this summer). 

Therefore, it is reasonable that we use the majority of the training set to build the sub-regressor, and
apply the rest small fraction for the blind test. This process can be applied in other machine-learning regression,
when there isn't additional data for a blind test.



\begin{acknowledgements}

This work was supported by the National Program on Key Research and Development
Project (Grant No. 2016YFA0400804) and
the National Natural Science Foundation of China (NSFC)
through grants NSFC-11603038/11425313/11403056.
This work presents results from the European Space Agency (ESA) space mission Gaia. Gaia data
are being processed by the Gaia Data Processing and Analysis Consortium (DPAC).
Funding for the DPAC is provided by national institutions, in particular the institutions
participating in the Gaia MultiLateral Agreement (MLA). The Gaia mission website is
\url{https://www.cosmos.esa.int/gaia}. The Gaia archive website is \url{https://archives.esac.esa.int/gaia}.

The Guoshoujing Telescope (the Large Sky Area Multi-Object
Fiber Spectroscopic Telescope, LAMOST) is a National Major
Scientific Project which is built by the Chinese Academy of
Sciences, funded by the National Development and Reform Commission,
and operated and managed by the National Astronomical Observatories,
Chinese Academy of Sciences.

Funding for the Sloan Digital Sky Survey IV has been provided by the Alfred P.
Sloan Foundation, the U.S. Department of Energy Office of Science, and the
Participating Institutions. SDSS-IV acknowledges
support and resources from the Center for High-Performance Computing at
the University of Utah. The SDSS web site is \url{http://www.sdss.org/}.

SDSS-IV is managed by the Astrophysical Research Consortium for the
Participating Institutions of the SDSS Collaboration including the
Brazilian Participation Group, the Carnegie Institution for Science,
Carnegie Mellon University, the Chilean Participation Group, the French
Participation Group, Harvard-Smithsonian Center for Astrophysics,
Instituto de Astrof\'isica de Canarias, The Johns Hopkins University,
Kavli Institute for the Physics and Mathematics of the Universe (IPMU) /
University of Tokyo, Lawrence Berkeley National Laboratory,
Leibniz Institut f\"ur Astrophysik Potsdam (AIP),
Max-Planck-Institut f\"ur Astronomie (MPIA Heidelberg),
Max-Planck-Institut f\"ur Astrophysik (MPA Garching),
Max-Planck-Institut f\"ur Extraterrestrische Physik (MPE),
National Astronomical Observatories of China, New Mexico State University,
New York University, University of Notre Dame,
Observat\'ario Nacional / MCTI, The Ohio State University,
Pennsylvania State University, Shanghai Astronomical Observatory,
United Kingdom Participation Group,
Universidad Nacional Aut\'onoma de M\'exico, University of Arizona,
University of Colorado Boulder, University of Oxford, University of Portsmouth,
University of Utah, University of Virginia, University of Washington, University of Wisconsin,
Vanderbilt University, and Yale University.
\end{acknowledgements}


\begin{thebibliography}{99}
%
%
%
\bibitem[Allende Prieto et al.(2008)]{Allende08} Allende Prieto, C., Sivarani, T., Beers, T.~C., et al.\ 2008, \aj, 136, 2070

\bibitem[Andrae et al.(2018)]{Andrae18} Andrae, R., Fouesneau, M., Creevey, O., et al.\ 2018, \aap, 616, A8
%
%
\bibitem[Bai et al.(2018)]{Bai18} Bai, Y., Liu, J.-F., \& Wang, S.\ 2018, Research in Astronomy and Astrophysics, 18, 118

\bibitem[Bai et al.(2019)]{Bai19} Bai, Y., Liu, J.-F., Wang, S., \& Yang, F.\ 2019, \aj, 157, 9
%
\bibitem[Bailer-Jones et al.(2013)]{Bailer13} Bailer-Jones, C.~A.~L., Andrae, R., Arcay, B., et al.\ 2013, \aap, 559, A74

\bibitem[Bailer-Jones et al.(2018)]{Bailer18} Bailer-Jones, C.~A.~L., Rybizki, J., Fouesneau, M., Mantelet, G., \& Andrae, R.\ 2018, \aj, 156, 58
%
%

\bibitem[Banko \& Brill (2001)]{Banko01} Banko, M. \& Brill, E. 2001, Scaling to very very large corpora for natural language disambiguation.
  In Proceedings of ACL-2001, pages 26¨C33. 
%
%
%
%
\bibitem[Bijaoui et al.(2012)]{Bijaoui12} Bijaoui, A., Recio-Blanco, A., de Laverny, P., \& Ordenovic, C. 2012, StMet, 9, 55
%
%
%
\bibitem[Breiman (2001)]{Breiman01} Breiman, L. 2001, in Random Forests, 45, pp 5-32
%
%
%
%
%

\bibitem[Davenport et al.(2014)]{Davenport14} Davenport, J.~R.~A., Ivezi{\'c}, {\v Z}., Becker, A.~C., et al.\ 2014, \mnras, 440, 3430
%

\bibitem[Du et al.(2012)]{Du12} Du, B., Luo, A., Zhang, J., Wu, Y., \& Wang, F.\ 2012, \procspie, 8451, 845137

%
%
\bibitem[Gaia Collaboration et al.(2016)]{Gaia16} Gaia Collaboration, Prusti, T., de Bruijne, J.~H.~J., et al.\ 2016, \aap, 595, A1
%
\bibitem[Gaia Collaboration et al.(2018)]{Gaia18} Gaia Collaboration, Brown, A.~G.~A., Vallenari, A., et al.\ 2018, \aap, 616, A1

\bibitem[Gao et al.(2015)]{Gao15} Gao, H., Zhang, H.-W., Xiang, M.-S., et al.\ 2015, Research in Astronomy and Astrophysics, 15, 220

%
%
\bibitem[Garc{\'{\i}}a P{\'e}rez et al.(2016)]{Garcia16} Garc{\'{\i}}a P{\'e}rez, A.~E., Allende Prieto, C., Holtzman, J.~A., et al.\ 2016, \aj, 151, 144
%
%

\bibitem[Green et al.(2018)]{Green18} Green, G.~M., Schlafly, E.~F., Finkbeiner, D., et al.\ 2018, \mnras, 478, 651
%
%
%
%
%
%
%
%
%
%

\bibitem[Koleva et al.(2009)]{Koleva09} Koleva, M., Prugniel, P., Bouchard, A., \& Wu, Y.\ 2009, \aap, 501, 1269

%
%
\bibitem[Kunder et al.(2017)]{Kunder17} Kunder, A., Kordopatis, G., Steinmetz, M., et al.\ 2017, \aj, 153, 75
%
%
\bibitem[Lee et al.(2008a)]{Lee08a} Lee, Y.~S., Beers, T.~C., Sivarani, T., et al.\ 2008a, \aj, 136, 2022
%
\bibitem[Lee et al.(2008b)]{Lee08b} Lee, Y.~S., Beers, T.~C., Sivarani, T., et al.\ 2008b, \aj, 136, 2050
%
\bibitem[Lee et al.(2015)]{Lee15} Lee, Y.~S., Beers, T.~C., Carlin, J.~L., et al.\ 2015, \aj, 150, 187
%
%
%
%
%
%
\bibitem[Luo et al.(2015)]{Luo15} Luo, A.-L., Zhao, Y.-H., Zhao, G., et al.\ 2015, Research in Astronomy and Astrophysics, 15, 1095

\bibitem[Luri et al.(2018)]{Luri18} Luri, X., Brown, A.~G.~A., Sarro, L.~M., et al.\ 2018, \aap, 616, A9
%
\bibitem[Majewski et al.(2017)]{Majewski17} Majewski, S.~R., Schiavon, R.~P., Frinchaboy, P.~M., et al.\ 2017, \aj, 154, 94

\bibitem[Mathur et al.(2017)]{Mathur17} Mathur, S., Huber, D., Batalha, N.~M., et al.\ 2017, \apjs, 229, 30
%
%
%
\bibitem[M{\'e}sz{\'a}ros et al.(2013)]{Meszaros13} M{\'e}sz{\'a}ros, S., Holtzman, J., Garc{\'{\i}}a P{\'e}rez, A.~E., et al.\ 2013, \aj, 146, 133
%
%
%
%
%

\bibitem[Pelisoli et al.(2019)]{Pelisoli19} Pelisoli, I., Bell, K.~J., Kepler, S.~O., \& Koester, D.\ 2019, \mnras, 482, 3831

\bibitem[Pil{\'a}szy \& Tikk (2009)]{Pilaszy09} Pil{\'a}szy, I. \& Tikk D. 2006, Recommending new movies: even a few ratings
  are more valuable than metadata, Proceedings of the third ACM conference on Recommender systems, October 23-25, New York, New York, USA

\bibitem[Recio-Blanco et al.(2006)]{Recio06} Recio-Blanco, A., Bijaoui, A., \& de Laverny, P.\ 2006, \mnras, 370, 141
%

\bibitem[Sahlholdt et al.(2019)]{Sahlholdt19} Sahlholdt, C.~L., Feltzing, S., Lindegren, L., \& Church, R.~P.\ 2019, \mnras, 482, 895


\bibitem[Sch{\"o}nrich et al.(2012)]{Schonrich12} Sch{\"o}nrich, R., Binney, J., \& Asplund, M.\ 2012, \mnras, 420, 1281

\bibitem[Smolinski et al.(2011)]{Smolinski11} Smolinski, J.~P., Lee, Y.~S., Beers, T.~C., et al.\ 2011, \aj, 141, 89
%
%
%
\bibitem[Steinmetz et al.(2006)]{Steinmetz06} Steinmetz, M., Zwitter, T., Siebert, A., et al.\ 2006, \aj, 132, 1645
%
%
%
%
%
%
\bibitem[Wang et al.(2016)]{Wang16} Wang, J., Shi, J., Zhao, Y., et al.\ 2016, \mnras, 456, 672

\bibitem[Wang et al.(2009a)]{Wang09a}Wang, S., Tang, K., \& Yao, X. 2009, Proc. Int. Joint Conf. Neural Netw. pp. 3259-3266 

\bibitem[Wang et al.(2009b)]{Wang09b}Wang, S. \& Yao, X. 2009, Proc. IEEE Symp. Computat. Intell. Data Mining pp. 324-331 

\bibitem[Wu et al.(2014)]{Wu14} Wu, Y., Du, B., Luo, A.L., et al. 2014, IAUS, 306, 340
%
%
%
\bibitem[Yanny et al.(2009)]{Yanny09} Yanny, B., Rockosi, C., Newberg, H.~J., et al.\ 2009, \aj, 137, 4377

%
%

\end{thebibliography}
\end{document}